\documentclass[a4paper,11pt]{article}
\input{paper.sty}

\title{Searching for spin-2 ULDM with gravitational waves interferometers}

\author[a]{Juan Manuel Armaleo,}
\author[a]{Diana L\'opez Nacir,}
\author[b]{and Federico R.~Urban}
\affiliation[a]{Departamento de Física  Juan Jos\'e Giambiagi, FCEyN UBA and IFIBA CONICET-UBA,
 Facultad de Ciencias Exactas y Naturales\\Ciudad Universitaria, Pabellon I, 1428 Buenos Aires, Argentina}
\affiliation[b]{CEICO, FZU--Institute of Physics of the Czech Academy of Sciences\\Na Slovance 2, 182 21 Praha 8, Czech Republic}

\abstract{The detection of gravitational waves from merging binaries has ushered in the era of gravitational wave interferometer astronomy.  Besides these strong, transient, calamitous events, much weaker signals can be detected if the oscillations are nearly monochromatic and ``continuous'', that is, coherent over a long time.  
In this work we show that ultra-light dark matter of spin two, owing to its universal coupling \(\al\) to Standard Model fields, generates a signal that is akin to but distinct from a continuous gravitational wave.  We show that this signal could be detected with current and planned gravitational wave interferometers.  In the event of a null detection, current facilities could constrain the coupling to be below \(\al\sim10^{-7}\) for frequencies of tens of Hz, corresponding to dark matter masses around the \(10^{-13}\)~eV mark.  Future facilities could further lower these upper limits and extend them to smaller masses down to \(10^{-18}\)~eV.  These limits would be the most stringent bounds on the spin-2 Yukawa fifth force strength, parametrised by \(\al\), in the frequency ranges accessible by gravitational wave interferometers.  The implementation of this type of searches for gravitational wave interferometers would therefore further our grasp of both dark matter and gravity.}

\begin{document}
\maketitle
\flushbottom


\section{Introduction}
\label{sec:intro}

The recent detection of gravitational waves (GWs) has consecrated gravitational wave interferometers (GWIs) as a vital experimental tool for astronomy, cosmology, and fundamental physics~\cite{LIGOScientific:2018mvr,Abbott:2020niy}.  The GWs detected thus far are a product of cataclysmic transient events, such as binary black hole mergers.  These signals are strong, with gravitational strain of the order of \(h\sim10^{-21}\), but very short, from a fraction of a second to several seconds.  Much weaker signals can be detected if they are coherent over a longer time, such as the continuous GWs (CWs) emitted by rapidly spinning neutron stars~\cite{Riles:2017evm} or ultra-compact Galactic binaries~\cite{Nelemans:2001hp}.  In the former case, recent searches for this type of signal have been performed in~\cite{Pisarski:2019vxw,Dergachev:2020fli,Steltner:2020hfd}.  Having detected no CWs, this set the upper limit \(h\sim10^{-25}\) on the maximum strain for this type of signal at frequencies of about \(f\sim10^2\)~Hz.

Another important source of CWs is due to the scattering of ultra-light bosons off black holes via a mechanism known as superradiance~\cite{Brito_2020}.  Bosons with masses \(m\ll1\)~eV, are predicted in theories beyond the Standard Model of particle physics, and are an excellent candidate for the cosmological dark matter dubbed ultra-light dark matter (ULDM)~\cite{Preskill:1982cy,Abbott:1982af,Dine:1982ah,Turner:1983he,Nelson:2011sf,Ferreira:2020fam}.  In particular, spin-2 ULDM is especially interesting because it arises as a modification of gravity itself, even though it is in the guise of an additional particle, the dark matter~\cite{Marzola:2017,Aoki:2017cnz}.  Searches for CWs produced by superradiance have been carried out only for spin-0 bosons~\cite{Palomba:2019vxe,Ng:2020ruv}, whereas no limits on the properties of spin-2 ultra-light bosons with GWI data exist yet.

In this work we show that, if ULDM has spin two, it interacts with GWIs in a way that, owing to its quasi-monochromaticity and persistence, closely resembles CWs.  The spin-2 ULDM-CW signal can be detected by existing Earth-based facilities such as advanced LIGO~\cite{TheLIGOScientific:2014jea} / advanced Virgo~\cite{TheVirgo:2014hva} (HLV) in their entire accessible frequency range, approximately corresponding to masses \(4\times10^{-14}~\mathrm{eV}\lesssim m \lesssim4\times10^{-11}~\mathrm{eV}\).  Furthermore, planned facilities such as LISA~\cite{Baker:2019nia}, DECIGO~\cite{Seto:2001qf}, and the BBO~\cite{Harry:2006fi} will extend this range down to \(m\sim\mathrm{few}\times10^{-19}~\mathrm{eV}\).  The spin-2 ULDM-CW signal is produced by the coherent oscillations of the ULDM field which is universally coupled to Standard Model fields and is unrelated to superradiance; this is similar to dark photon dark matter, where the ULDM carries additional interactions~\cite{Pierce:2018xmy,Miller:2020vsl}\footnote{Other types of direct interactions between ULDM and matter have also been considered, see, e.g., \cite{Arvanitaki:2014faa,Morisaki:2018htj,Grote:2019uvn,Michimura:2020vxn}} (notice however that in the spin-2 case the interaction can not be tuned away).  Moreover, regardless the spin, if the ULDM field only interacts gravitationally, the signal is undetectable by GWIs~\cite{Aoki:2016kwl}.  Our findings demonstrate that, in case of a null result, GWIs can place some of the most stringent bounds on the spin-2 Yukawa fifth force strength \(\al\) in the frequency ranges accessible by GWIs.

This paper is structured as follows.  In Section~\ref{sec:maths} we compute the strength and shape of the expected signal from spin-2 ULDM for the frequency ranges of interest for GWIs.  In Section~\ref{sec:res} we present our results, and in Section~\ref{sec:out} we put them into context and give an outlook for future work.  We work with units in which \(c=k_B=\hslash=1\), and use latin indices \((i,j,\ldots)\in[1,3]\) for spatial tensor components.

\section{The shape and strength of the signal}
\label{sec:maths}

The behaviour of the spin-2 ULDM in sufficiently small regions inside the local dark matter halo is described by the oscillating tensor field~\cite{Marzola:2017}
\begin{align}
    \Mij(t) &= \frac{\sqrt{2\rhoDM}}{m}\cos{(mt+\Upsilon)}\epij(\vr) \,, \label{eq:Mij}
\end{align}
where \(\rhoDM\) is the observed local dark matter energy density, for which we assume   \(\rhoDM=0.3\)~GeV/cm\(^3\)~\cite{Piffl:2014mfa,Evans:2018bqy,2015ApJ...814...13M}, and \(\Upsilon\) is a random phase.  The five polarisations of the spin-2 field are encoded in the \(\epij(\vr)\) tensor, which has unit norm and zero trace, is symmetric and is direction-dependent via the unit vector \(\vr\)~\cite{Maggiore:1900zz}.  The solution \Eq{eq:Mij} assumes a single frequency \(2\pi f=m\) and a coherent polarisation structure.  The latter is justified for scales shorter than the characteristic scale of the inhomogeneities of the ULDM field, which is given by the de~Broglie wavelength \(\ldB \deq 2\pi/mv = 1/fv\) where \(v\sim 10^{-3}\) is the effective velocity of the ULDM.  Thus, owing to the fact that \(\ldB\) is much larger than the physical size of the GWIs and the distance between the HLV sites, we can safely neglect gradients (see~\cite{Armaleo:2020yml} for further discussion).  The coherence of the oscillation frequency is instead guaranteed up to a coherence time that is given by\footnote{Notice the definition of the coherence time differs from the one that is commonly used in the ULDM literature by a factor~4.  We adopt the definition used in the GW literature here.} \(\tcoh \deq 4\pi/mv^2 = 2/fv^2\).  Given that a typical GWI observation run will last for much longer than \(\tcoh\), a more precise description of the ULDM field would be a superposition of plane waves, see~\cite{Pierce:2018xmy,Miller:2020vsl}; we neglect this for our order of magnitude estimates\footnote{This solution is also valid provided that the energy, or frequency, scales of the system is well below the ultra-violet cutoff of the effective field theory, that is \(f \sim m/2\pi \ll (M_\text{P}m^2)^{1/3}\)~\cite{Akrami:2015qga}; this is easily verified for all the values of the spin-2 mass \(m\) we consider in this work.}.

In the ULDM reference frame \((\vp,\vq,\vr)\) the polarisations of the spin-2 field can be described as \(\epij(\vr) \deq \sum_\kappa \vep_\kappa {\cal Y}^\kappa_{ij}(\vr)\)~\cite{Maggiore:1900zz,Armaleo:2019gil}, where the summation runs over the five amplitudes \(\left\{ \vepCross, \vepPlus, \vepL,\vepR, \vepS \right\}\) that obey \(\sum_\kappa \vep_\kappa^2=1\)---the overall amplitude is fixed by the requirement that \(\Mij\) makes up all of the dark matter.  The five polarisation matrices are given by
\begin{align}
    {\cal Y}^\times_{ij} &\deq \frac{1}{\sqrt2} \left(p_i q_j + q_i p_j\right) \,, & {\cal Y}^+_{ij} &\deq \frac{1}{\sqrt2} \left(p_i p_j - q_i q_j\right) \,, & \nn\\
    {\cal Y}^L_{ij} &\deq \frac{1}{\sqrt2} \left(q_i r_j + r_i q_j\right) \,, & {\cal Y}^R_{ij} &\deq \frac{1}{\sqrt2} \left(p_i r_j + r_i p_j\right) \,, & \nn\\
    {\cal Y}^S_{ij} &\deq \frac{1}{\sqrt6} \left(3 r_i r_j - \delta_{ij}\right) \,. &&& \nn
\end{align}
Notice that, unlike for CWs, there is no propagation along the \(\vr\) direction, which in our case serves merely as reference for the decomposition in tensor, vector, and scalar helicities according to their behaviour under a rotation about \(\vr\) (see also Appendix~\ref{app:not}).
 
Spin-2 ULDM couples to Standard Model fields \(\Psi\) as~\cite{Marzola:2017}
\begin{align}\label{eq:int}
    S_\text{int}[g,\Mij,\Psi] & \deq -\frac{\al}{2\mpl} \int\!\dd^4x\,\sqrt{-g} \Mij T_\Psi^{ij} \,,
\end{align}
where \(T_\Psi^{ij}\) is the stress tensor of the fields \(\Psi\) and \(\mpl\) is the reduced Planck mass.  At leading (linear) order in \(\alpha\) the interaction \Eq{eq:int} can be absorbed into a redefinition of the metric \(g_{ij}\to g_{ij} + \al  M_{ij}/\mpl\)~\cite{Armaleo:2020yml}.  Therefore, the effect of spin-2 ULDM on the detector can be equivalently described by the gravitational effect of an oscillating metric perturbation \(\hij\) given by
\begin{align}
    \hij(t) &= \frac{\al}{\mpl}\Mij(t) = \frac{\al\sqrt{2\rhoDM}}{m\mpl}\cos{(mt+\Upsilon)}\epij(\vx) \,.
\end{align}
The parameter \(\al\) is idiosyncratic for spin-2 ULDM because it is required by the self-consistency of the model, such as in bigravity~\cite{Babichev:2016bxi}.  This parameter defines the inverse ULDM self-interaction strength: there is no ULDM at all with \(\al\rar0\) because the ULDM field becomes infinitely strongly coupled in this limit.  Furthermore, spin-2 ULDM is ineluctably coupled universally to standard matter fields, so that ULDM will appear as a Yukawa-like fifth force modification of the gravitational potential \(\Phi\) in the weak field regime, for which \(\al\) quantifies the strength: \(\Phi\rar\Phi\left[1+\al^2\exp(-mr)\right]\).  The strength of this fifth force for different values of the mass \(m\) (or, equivalently, frequency) is constrained by several experiments and tests of gravity~\cite{Murata:2014nra,Sereno:2006mw}: we call this maximal coupling \(\al=\al_Y\).

In the reference frame of the detector, \((\vx,\vy,\vz)\), the response function \(D^{ij}\) is given by the differential change in the length of the detector arms directed along the unit vectors \(\vn\) and \(\vm\) as \(D^{ij} = (n^i n^j - m^i m^j)/2\)~\cite{Maggiore:1900zz}.  The signal is the combination of the variation of the metric perturbation and the response function:
\begin{align}
    h(t) &\deq D^{ij} h_{ij}(t) = \frac{\al\sqrt{\rhoDM}}{\sqrt2 m\mpl}\cos{(mt+\Upsilon)}\Delta\vep \deq  h_s\sin{(mt)} + h_c\cos{(mt)}\,, \label{eq:signal}
\end{align}
where we defined \(\Delta\vep \deq \,\epij (n^i n^j - m^i m^j)\), and introduced the sine \(h_{s}\) and cosine \(h_{c}\)     amplitudes.  This is the central equation of the paper.

The theoretical spin-2 ULDM-CW signal \Eq{eq:signal} presents two key features.  First, the signal is inversely proportional to the spin-2 boson mass \(m\).  This inverse linear scaling is also found in dark photon dark matter, where the spin-1 ULDM field carries a additional charges such as baryon number \(B\) or baryon minus lepton number \(B-L\), through which the ULDM directly interacts with the mirrors of the detector~\cite{Pierce:2018xmy,Miller:2020vsl}.  The inverse linear dependence should be compared with the generic inverse \emph{quadratic} dependence obtained by pure gravitational interaction~\cite{Aoki:2016kwl}.  In other words, in absence of non-gravitational interactions, the signal strength decays much more rapidly with increasing mass (or frequency).  This makes it practically impossible to detect such a signal with future GWIs, let alone existing ones.  Second, the spin-2 ULDM-CW signal has a unique geometric structure that sets it apart from other CWs.  Explicitly we have
\begin{align}
    \Delta\vep  &= \sqrt{2} \vepCross\left[\left(\vp\cdot\vn\right) \left(\vq\cdot\vn\right) - \left(\vp\cdot\vm\right) \left(\vq\cdot\vm\right)\right] + \frac{\vepPlus}{\sqrt{2}}\left[\left(\vp\cdot\vn\right)^2 - \left(\vq\cdot\vn\right)^2 -\left(\vp\cdot\vm\right)^2 + \left(\vq\cdot\vm\right)^2\right] \nn\\
                &~~+ \sqrt{2} \vepL\left[\left(\vq\cdot\vn\right) \left(\vr\cdot\vn\right) - \left(\vq\cdot\vm\right) \left(\vr\cdot\vm\right)\right] + \sqrt{2}\vepR\left[\left(\vp\cdot\vn\right) \left(\vr\cdot\vn\right) - \left(\vp\cdot\vm\right) \left(\vr\cdot\vm\right)\right] \nn\\
                &~~+ \sqrt{\frac{3}{2}}\,\vepS\left[\left(\vr\cdot\vn\right)^2 - \left(\vr\cdot\vm\right)^2\right] \label{eq:signal_rpq} \\
                &=\frac{\cos2\phi}{\sqrt{2}}  \left[\vepPlus\left(\cos^2\theta+1\right) + \vepR\,\sin2\theta + \sqrt3\,\vepS\,\sin^2\theta\right] -\sqrt{2}\sin2\phi\left(\vepCross\,\cos\theta + \vepL\,\sin\theta\right) \,, \label{eq:signal_xy}
\end{align}
where in obtaining the last expression we have set \(\vn=\vx\) and \(\vm=\vy\), which we can always do for a single L-shaped detector, and we have defined the ULDM reference frame in terms of the detector's frame as \(\vr = (\sin\theta\cos\phi,\sin\theta\sin\phi,\cos\theta)\), \(\vp = (\cos\theta\cos\phi,\cos\theta\sin\phi,-\sin\theta)\), \(\vq = (-\sin\phi,\cos\phi,0)\); the origins of the two frames are connected by the vector \(r\vr\).

Before moving on to our results, a comment is in order here.  The detector is moving with respect to the ULDM, this motion being the result of three contributions: (1) the Earth is rotating about its axis with equatorial velocity of approximately \(v\sim10^{-6}\) (this only applies to Earth-bound detectors); (2) the Earth is moving along its orbit around the Sun with speed \(v\sim10^{-4}\); (3) the Solar System is moving through the dark matter halo at a speed of \(v\sim10^{-3}\) causing what is known as the dark matter wind.  Therefore, in principle we should Lorentz-boost the ULDM frame to the reference frame of the detector.  However, owing to the smallness of the velocities in question, the effect of the boost on \(r\vr\) amounts to less than a percent correction to the theoretical signal and can be safely neglected.  The relative acceleration of the two frames also induces a Doppler frequency shift \(\Delta f_\text{Doppler}\) that affects the spin-2 ULDM-CW signal, and that needs to be accounted for when designing a data analysis pipeline~\cite{Miller:2020vsl,Frasca:2005ey,DAntonio:2018sff}.

All-sky searches for CWs with Earth-bound GWIs resort to semi-coherent methods because it is not computationally feasible to analyse the data from the entire observation campaign in a fully coherent way\footnote{In the case of space-based detectors such as the upcoming LISA interferometer, owing to the sparse sampling frequency of around 1~Hz, compared to the HLV sampling of about \(10^4\)~Hz, this is not an issue.}~\cite{Brady:1998nj,Krishnan:2004sv,Antonucci:2008jp,PhysRevD.90.042002}.  In semi-coherent methods the whole data set is broken into shorter time chunks of length \(\tchunk\), each of which is then analysed coherently but separately.  One of the advantages of this approach is that, by choosing \(\tchunk<\tdop\deq1/\Delta f_\text{Doppler}\) the Doppler frequency shift can be neglected\footnote{To be more precise, within each chunk the instantaneous Doppler shift that would contribute to \(\dot{f}\) can be neglected, i.e., the frequency is held constant.  Nevertheless, in CW searches, in order to identify viable source candidates for the follow-up steps in the hierarchical semi-coherent analysis, the predicted Doppler shift for each chunk and each location in the sky needs to be corrected for.  Being there no ``sky location'' for ULDM searches, this is not a concern.}.  Moreover, one should ensure that \(\tchunk<\tcoh\) in order to have a stable ULDM configuration within a given chunk.  The sensitivity for a coherent analysis over the whole observation campaign time \(\tobs\) scales as \(\tobs^{-1/2}\).  In semi-coherent methods, assuming that all \(N\) chunks last the same time \(\tchunk\) and all together they cover the whole observation run such that \(\tobs=N\tchunk\), the sensitivity scales instead as \(N^{-1/4}\tchunk^{-1/2} = \tobs^{-1/4}\tchunk^{-1/4}\).  Thanks to the coherence of the signal, even within the limitation of the semi-coherent methods, the actual sensitivity attained by the HLV collaboration for CW searches is more than a factor \(10^{-3}\) smaller than the design sensitivity \(h_0\) for transient events~\cite{Pisarski:2019vxw,Dergachev:2020fli,Steltner:2020hfd}.

The semi-coherent techniques have been adapted and optimised, taking into account the coherence time and the geometry of the signal, for dark photon dark matter searches~\cite{Miller:2020vsl}.  They can therefore be tailored for spin-2 ULDM-CW searches by replacing the average over the different polarisations of ULDM waves (which for the spin 1 dark photon case amounts to a factor \(\sqrt{2}/3\)~\cite{Pierce:2018xmy,Miller:2020vsl}) with \(\sqrt{\langle\Delta\varepsilon^2\rangle} = \sqrt{2/5}\).  We define the effective theoretical strain amplitude \(h\) for the spin-2 ULDM-CW signal as the root mean square average, taken over all the polarisation angles and the random phase \(\Upsilon\), of the sine and cosine amplitudes of \Eq{eq:signal}:
\begin{align}\label{eq:signalaver}
    h &\deq \langle h_s^2+h_c^2\rangle^{1/2} = \frac{\al\sqrt{\rhoDM}}{\sqrt{5}m\mpl} \,.
\end{align}
 
\section{Results}
\label{sec:res}

\begin{figure}[htbp]
\centering
	\includegraphics[width=1.0\textwidth]{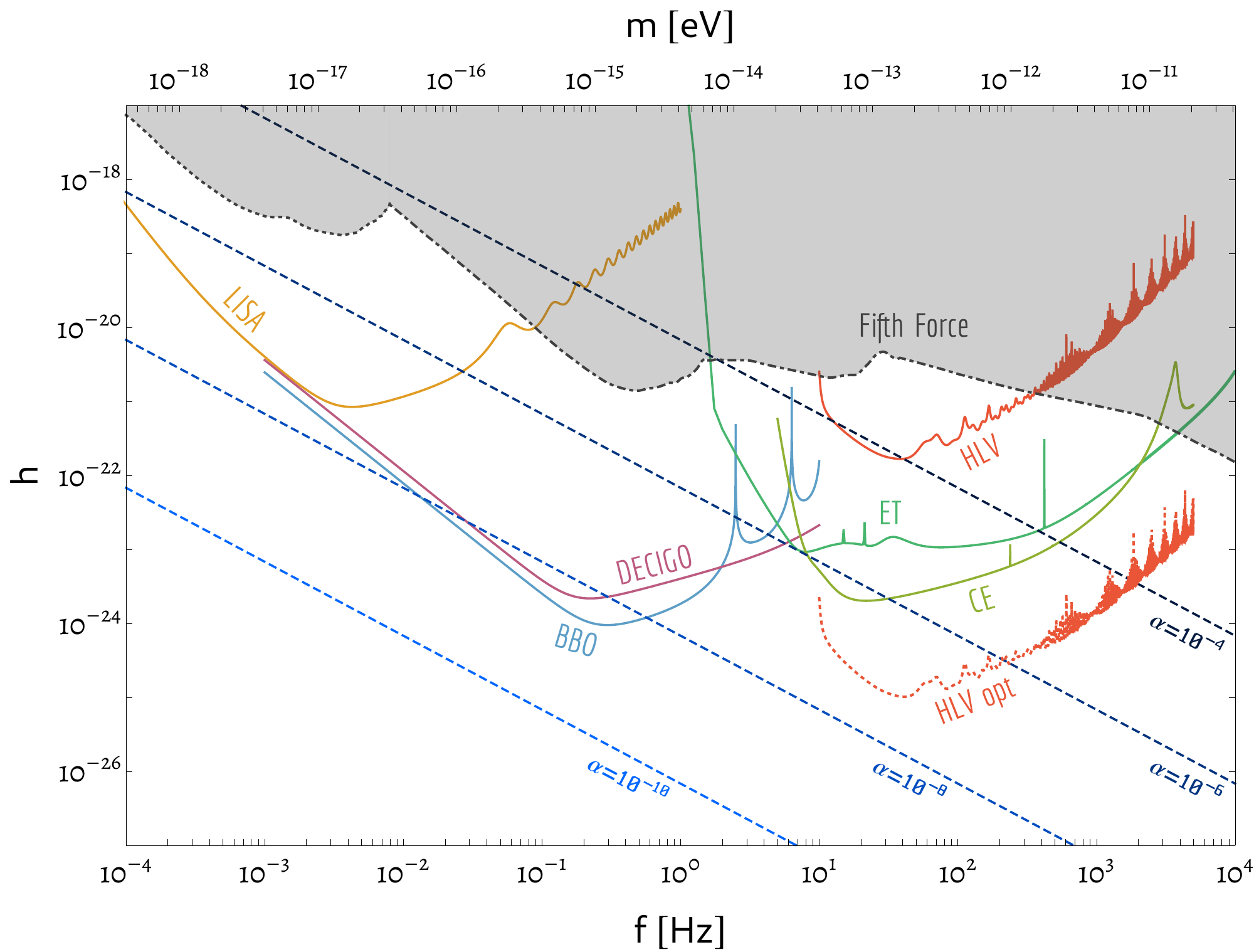}
	\caption{Design sensitivity \(h=h_0\) for several current and planned GWIs, as a function of frequency (solid lines).  The dotted line ``HLV opt'' is the optimised sensitivity obtained with a semi-coherent method tailored for spin-2 ULDM-CW searches (Appendix \ref{app:opt}).  Overlaid as dashed lines are the signal strains \(h\) of \Eq{eq:signalaver} for different values of the parameter \(10^{-4}\leq\al\leq10^{-10}\).  The dot-dashed black line is the spin-2 ULDM-CW strain corresponding to the maximal values of \(\al\) allowed by fifth force constraints, \(h=h(\al_Y)\) with \(\al_Y\) obtained from~\cite{Murata:2014nra,Sereno:2006mw}; the region above this line is excluded.}\label{fig:signal}
\end{figure}

In order to estimate the values of \(\al\) accessible with GWIs, we compare the expected theoretical signal \(h\) of \Eq{eq:signalaver} with the design sensitivities of a number of current and planned GWIs (Fig.~\ref{fig:signal}).  We find that the HLV detectors can nominally detect spin-2 ULDM for \(\al\gtrsim10^{-4}\) depending on the frequency (Fig.~\ref{fig:signal}).  We expect that a dedicated semi-coherent search for the spin-2 ULDM-CW signal will improve the range of detectable \(\al\) by a few orders of magnitude, potentially down to \(\al\sim10^{-7}\) or less for frequencies of tens of Hz, corresponding to masses around the \(10^{-13}\)~eV mark; this is shown in Fig.~\ref{fig:signal} as the dotted line ``HLV opt''---the details on how we obtained this curve can be found in Appendix~\ref{app:opt}.  In this frequency range, from \(f\sim10\)~Hz (\(m\sim4\times10^{-14}\)~eV) to \(f\sim10^3\)~Hz (\(m\sim4\times10^{-12}\)~eV) and beyond the planned experiments Einstein Telescope (ET)~\cite{Hild:2010id} and Cosmic Explorer (CE)~\cite{Evans:2016mbw} should reach sensitivities of order \(h_0\sim10^{-22}\text{---}10^{-23}\), further improving the chances to detect spin-2 ULDM.

Future facilities will also be able to probe much lower values of the ULDM mass.  In the intermediate frequency range \(0.1~\mathrm{Hz}\lesssim f \lesssim1\)~Hz, corresponding to \(4\times10^{-16}~\mathrm{eV}\lesssim m \lesssim4\times10^{-15}\)~eV, the BBO and DECIGO detectors are expected to attain sensitivities of order of \(h_0\sim10^{-23}\text{---}10^{-24}\)~\cite{Seto:2001qf,Harry:2006fi}.  This means these GWIs could detect a spin-2 ULDM-CW signal for \(\al\lesssim10^{-8}\) at those frequencies.  In the low frequency range the planned space-based interferometer LISA will reach a sensitivity of \(h_0\sim10^{-21}\) for \(f\sim10^{-2}\)~Hz (\(m\sim4\times10^{-17}\)~eV), which means that it could detect spin-2 ULDM with \(\al\sim10^{-7}\) and below.  These limits would be much improved with a dedicated pipeline for these interferometers, as is the case for HLV.  We collect all the sensitivities as compiled in~\cite{Schmitz:2020syl} and compare them to the theoretical signal in Fig.~\ref{fig:signal}---notice that strictly speaking these sensitivities are valid only for the standard tensor modes of GWs, namely the \(\vepCross\) and \(\vepPlus\) in our notation, but the differences are small and not relevant for our order of magnitude estimates~\cite{Zhang:2019oet}.

\section{Conclusion and outlook}
\label{sec:out}

GWIs are a unique tool to understand the nature of gravity.  In this work we have shown that GWIs have the potential to test the properties of gravity \emph{and} dark matter by detecting or constraining spin-2 ULDM.  In particular, we expect that the existing HLV facilities could detect spin-2 ULDM for values of the coupling parameter \(\al\) as little as \(\al\sim10^{-7}\) for frequencies of \(f\sim\mathrm{few}~\times100\)~Hz (that is, a Yukawa range \(\lambda\deq1/2\pi f\sim10^4\)~m).  A null result would place the most stringent limits the strength of the Yukawa-like fifth force modification of the inverse-square law of gravitational interaction, quantified by \(\al\), provided that the fifth force is carried by the dark matter.  Looking forward, future GWIs in the same frequency range can push this limit even further by up to two orders of magnitude, whereas planned facilities such as DECIGO and the BBO (\(f\sim0.1\)~Hz), and the milli-Hertz space-based LISA interferometer are expected to attain \(\al\lesssim10^{-7}\text{---}10^{-8}\) in their respective frequency ranges; limits which can be significantly improved with a dedicated pipeline for the spin-2 ULDM-CW signal.

Our results complement our previous studies on the bounds on the spin-2 ULDM coupling \(\al\) coming from PTAs~\cite{Armaleo:2020yml} and individual pulsar timing data~\cite{Armaleo:2019gil}, which cover the frequency range \(10^{-9}~\mathrm{Hz}\lesssim f \lesssim10^{-3}\)~Hz, and for which comparable limits on \(\al\) were obtained.

Our findings should be compared with existing limits on spin-2 ULDM coming from superradiance.  By measuring the spin and mass of known black holes and other astrophysical objects, the mass ranges  \(6.4\times10^{-22}~\mathrm{eV}\lesssim m \lesssim7.7\times10^{-21}~\mathrm{eV}\), \(1.8\times10^{-20}~\mathrm{eV}\lesssim m \lesssim1.8\times10^{-16}~\mathrm{eV}\) and \(2.2\times10^{-14}~\mathrm{eV}\lesssim m \lesssim2.8\times10^{-11}~\mathrm{eV}\)  are excluded, or else these black holes and celestial objects would not be there~\cite{Stott:2020gjj}.  These bounds are valid provided that any additional interactions that the bosons might possess are small enough not to interfere with the onset and development of superradiance.  In particular, these limits are valid if \(10^{-30}\,\mathrm{eV}/m\ll\al\ll1\)~\cite{Brito:2020lup}, which is verified for most of the parameter space we are considering.  Therefore, the limits we can obtain from GWIs will at the same time independently exclude some of the parameter space that is probed by superradiance, as well as test regions not accessible by it.  Spin-2 ULDM can also be detected thanks to the CW signal that superradiance produces, which is physically unrelated to and distinct from the signal we have described in this work~\cite{Brito:2020lup}; no such searches have been carried out yet for spin-2 ULDM.

In order to fully take advantage of GWI data to test spin-2 ULDM, a dedicated pipeline should be developed.  As we have shown in Sec.~\ref{sec:maths} the signal \Eq{eq:signal} has a peculiar geometric structure that is explicitly given in \Eq{eq:signal_rpq}.  Moreover, the ULDM signal is expected to be coherent for a time \(\tcoh = 2/fv^2\).  An optimised analysis molded onto the shape of this signal can not only improve the sensitivity of GWIs to spin-2 ULDM-CW, but also discriminate between ULDM and other sources of CWs at different frequencies, such as fast-spinning Galactic neutron stars (at high frequencies) or ultra-compact Galactic binaries (in the milli-Hertz band), CW coming from superradiance, and other variants of ULDM, furthering our grasp of dark matter and gravity.

\acknowledgments

The Authors would like to thank C.~Palomba for valuable discussion on the semi-coherent methods used in GW full-sky searches, R.~Brito and V.~Cardoso for an update on the status of spin-2 superradiance, and N.~Tamanini, A. Klein and R. Sturani for useful correspondence on CW searches with LISA.  FU is supported by the European Regional Development Fund (ESIF/ERDF) and the Czech Ministry of Education, Youth and Sports (MEYS) through Project CoGraDS - \verb|CZ.02.1.01/0.0/0.0/15_003/0000437|.  The work of DLN and JMA has been supported by CONICET, ANPCyT and UBA.

\appendix
\section{Notation}\label{app:not}

In order to make contact with our previous works~\cite{Armaleo:2019gil,Armaleo:2020yml} here we provide the dictionary that connects the two ways to express the spin-2 polarisations \(\epij(\vr) \deq \sum_\kappa \vep_\kappa {\cal Y}^\kappa_{ij}(\vr) \deq \sum_m a_m {\cal Y}^{2m}_{ij}(\vr)\):
\begin{align}
    {\cal Y}^\times_{ij} &\deq {\cal Y}^{2,-2}_{ij}\,, ~~~ {\cal Y}^+_{ij} \deq {\cal Y}^{2,2}_{ij}\,, ~~~ {\cal Y}^L_{ij} \deq {\cal Y}^{2,-1}_{ij}\,, ~~~ {\cal Y}^R_{ij} \deq {\cal Y}^{2,1}_{ij}\,, ~~~ {\cal Y}^S_{ij} \deq {\cal Y}^{2,0}_{ij} \,. \nn
\end{align}
The polarisation tensor is described by three amplitudes and two angles according to
\begin{align}\label{parampolariz}
    \vepCross\deq&\vepT\sin\chi\deq a_{-2}\deq\sin\eta\cos\beta\sin\chi  \,, & \vepPlus\deq&\vepT\cos\chi\deq a_{2}\deq\sin\eta\cos\beta\cos\chi \,, \nn\\
    \vepL\deq&\vepV\sin\tau\deq a_{-1}\deq\sin\eta\sin\beta\sin\tau \,, & \vepR\deq&\vepV\cos\tau\deq a_{1}\deq\sin\eta\sin\beta\cos\tau\,, \nn\\
    \vepS\deq&a_{0}\deq\cos\eta \,.&&
\end{align}
The amplitudes in this description obey \(\vepT^2+\vepV^2+\vepS^2=1\), or \(\sum_m a_m^2=1\).
 
The most generic symmetric polarisation tensor that is diffeomorphism-invariant can have up to six independent degrees of freedom~\cite{Lee:2008}.  This is indeed the case in many alternative theories of gravity.  In that case the most general decomposition of the polarisation tensor \(\epij(\vr)\) has two scalar modes: the breathing mode \(\vep_\mathrm{b}\) with \({\cal Y}^b_{ij} \propto \left(p_i p_j + q_i q_j\right)\) and the longitudinal mode \(\vep_\mathrm{l}\) with \({\cal Y}^b_{ij} \propto \left(r_i r_j\right)\).  These two modes, owing to the traceleness of our polarisation tensor \(\delta^{ij}\epij=0\), are combined into one single scalar mode \(\vepS\).

\section{Optimised sensitivity}\label{app:opt}

We provide here the details in the calculation of the optimised HLV sensitivity shown in Fig.~\ref{fig:signal}.  In order to adapt the theoretical sensitivity of the semi-coherent frequency-Hough method of~\cite{PhysRevD.90.042002}, originally introduced for CW searches with Earth-based detectors, one needs to take into account two factors, see~\cite{Miller:2020vsl}.  First, because the signal associated to the ULDM field is always present at the detector, rather than coming from a given direction in the sky, there is a factor of \(5/2\) that should be removed from Eq.~(67) of~\cite{PhysRevD.90.042002}---this factor comes from performing an average over sky directions.  Second, one needs to compute the average over the different polarisations of ULDM waves.  In the spin 1 dark photon dark matter case this is \(\sqrt{2}/3\)~\cite{Pierce:2018xmy,Miller:2020vsl}, whereas for our spin-2 waves we find \(\sqrt{\langle\Delta\varepsilon^2\rangle}= \sqrt{2/5}\) (we used Eq.~(\ref{eq:signal_xy}) and the parametrisation given in (\ref{parampolariz})).

The optimised sensitivity at a given confidence level \(\Gamma\) indicates the minimum signal amplitude which would produce a candidate detection in a fraction \(\geq\Gamma\) of a large number of repeated experiments.  This can be written as~\cite{Miller:2020vsl}
\begin{align}\label{minamplsen}
    h_{0,\rm opt} &\approx \frac{1.02}{N^{1/4} \theta_\text{thr}^{1/2}}\sqrt{\frac{S_n(f)}{T_\text{FFT,max}}}\left(\frac{p_0(1-p_0)}{p_1^2}\right)^{1/4}\sqrt{CR_\text{thr}-\sqrt{2}\text{erfc}^{-1}(2\Gamma)} \,, \nn\\
    N &= \frac{\tobs}{T_\text{FFT,max}} \,, \nn\\
    p_0 &= e^{-\theta_\text{thr}}-e^{-2\theta_\text{thr}}+\frac{1}{3}e^{-3\theta_\text{thr}} \,, \nn\\
    p_1 &= e^{-\theta_\text{thr}}-2e^{-2\theta_\text{thr}}+ e^{-3\theta_\text{thr}} \,. \nn
\end{align}
Here \(N\) is half the number of Fast Fourier Transforms (FFT) during the observation time \(\tobs\) (assuming the FFTs are interlaced by half), \(\theta_\text{thr}\) is the threshold for peak selection to create the so called {\it peakmap}, \(CR_\text{thr}\) is the threshold for candidate selection, \(S_n(f)\) is the noise power spectral density of the detector, and \(T_\text{FFT,max}\) is taken to be the maximum \(\tchunk\) given by the coherence time of the signal.  Following~\cite{Miller:2020vsl}, in obtaining the optimised sensitivity we use
\begin{align} 
    T_\text{FFT,max}\lessapprox\frac{2}{f}\frac{1}{v_\text{esc}^2}\approx \frac{6\times 10^5}{f}~\text{s} \,,
\end{align}
with \(v_\text{esc}\) the escape velocity of the DM in the local halo, and we set \(\theta_\text{thr}=2.5\),  \(CR_\text{thr}=5\) and \(\Gamma=0.95\).  Finally, we have used \(f S_n(f) = h_0(f)^2\) where \(h_0(f)\) for HLV can be found in~\cite{Schmitz:2020syl}, and \(\tobs= 1yr\).

\bibliographystyle{hieeetr}
\bibliography{biblio.bib}

\begin{thebibliography}{10}

\bibitem{LIGOScientific:2018mvr}
B.~Abbott {\em et~al.}, ``{GWTC-1: A Gravitational-Wave Transient Catalog of
  Compact Binary Mergers Observed by LIGO and Virgo during the First and Second
  Observing Runs},'' {\em Phys. Rev. X}, vol.~9, no.~3, p.~031040, 2019,
  1811.12907.

\bibitem{Abbott:2020niy}
R.~Abbott {\em et~al.}, ``{GWTC-2: Compact Binary Coalescences Observed by LIGO
  and Virgo During the First Half of the Third Observing Run},'' {\em preprint
  (arxiv: 2010.14527)}, Oct 2020, 2010.14527.

\bibitem{Riles:2017evm}
K.~Riles, ``{Recent searches for continuous gravitational waves},'' {\em Mod.
  Phys. Lett. A}, vol.~32, no.~39, p.~1730035, 2017, 1712.05897.

\bibitem{Nelemans:2001hp}
G.~Nelemans, L.~Yungelson, and S.~F. Portegies~Zwart, ``{The gravitational wave
  signal from the galactic disk population of binaries containing two compact
  objects},'' {\em Astron. Astrophys.}, vol.~375, pp.~890--898, 2001,
  astro-ph/0105221.

\bibitem{Pisarski:2019vxw}
B.~Abbott {\em et~al.}, ``{All-sky search for continuous gravitational waves
  from isolated neutron stars using Advanced LIGO O2 data},'' {\em Phys. Rev.
  D}, vol.~100, no.~2, p.~024004, 2019, 1903.01901.

\bibitem{Dergachev:2020fli}
V.~Dergachev and M.~A. Papa, ``{Results from the First All-Sky Search for
  Continuous Gravitational Waves from Small-Ellipticity Sources},'' {\em Phys.
  Rev. Lett.}, vol.~125, no.~17, p.~171101, 2020, 2004.08334.

\bibitem{Steltner:2020hfd}
B.~Steltner, M.~Papa, H.-B. Eggenstein, B.~Allen, V.~Dergachev, R.~Prix,
  B.~Machenschalk, S.~Walsh, S.~Zhu, and S.~Kwang, ``{Einstein@Home all-sky
  search for continuous gravitational waves in LIGO O2 public data},'' {\em
  preprint (arxiv: 2009.12260)}, Sep 2020, 2009.12260.

\bibitem{Brito_2020}
R.~Brito, V.~Cardoso, and P.~Pani, ``{Superradiance}: {New Frontiers in Black
  Hole Physics},'' {\em Lect. Notes Phys.}, vol.~971, Springer (2020),
  1501.06570.

\bibitem{Preskill:1982cy}
J.~Preskill, M.~B. Wise, and F.~Wilczek, ``{Cosmology of the Invisible
  Axion},'' {\em Phys. Lett.}, vol.~B120, pp.~127--132, 1983.

\bibitem{Abbott:1982af}
L.~F. Abbott and P.~Sikivie, ``{A Cosmological Bound on the Invisible Axion},''
  {\em Phys. Lett.}, vol.~B120, pp.~133--136, 1983.

\bibitem{Dine:1982ah}
M.~Dine and W.~Fischler, ``{The Not So Harmless Axion},'' {\em Phys. Lett.},
  vol.~B120, pp.~137--141, 1983.

\bibitem{Turner:1983he}
M.~S. Turner, ``{Coherent Scalar Field Oscillations in an Expanding
  Universe},'' {\em Phys. Rev.}, vol.~D28, p.~1243, 1983.

\bibitem{Nelson:2011sf}
A.~E. Nelson and J.~Scholtz, ``{Dark Light, Dark Matter and the Misalignment
  Mechanism},'' {\em Phys. Rev.}, vol.~D84, p.~103501, 2011, 1105.2812.

\bibitem{Ferreira:2020fam}
E.~G. Ferreira, ``{Ultra-Light Dark Matter},'' {\em preprint (arxiv:
  2005.03254)}, May 2020, 2005.03254.

\bibitem{Marzola:2017}
L.~Marzola, M.~Raidal, and F.~R. Urban, ``{Oscillating Spin-2 Dark Matter},''
  {\em Phys. Rev.}, vol.~D97, no.~2, p.~024010, 2018, 1708.04253.

\bibitem{Aoki:2017cnz}
K.~Aoki and K.-i. Maeda, ``{Condensate of Massive Graviton and Dark Matter},''
  {\em Phys. Rev.}, vol.~D97, no.~4, p.~044002, 2018, 1707.05003.

\bibitem{Palomba:2019vxe}
C.~Palomba {\em et~al.}, ``{Direct constraints on ultra-light boson mass from
  searches for continuous gravitational waves},'' {\em Phys. Rev. Lett.},
  vol.~123, p.~171101, 2019, 1909.08854.

\bibitem{Ng:2020ruv}
K.~K. Ng, S.~Vitale, O.~A. Hannuksela, and T.~G. Li, ``{Constraints on
  ultralight scalar bosons within black hole spin measurements from
  LIGO-Virgo's GWTC-2},'' {\em preprint (arxiv: 2011.06010)}, Nov 2020,
  2011.06010.

\bibitem{TheLIGOScientific:2014jea}
J.~Aasi {\em et~al.}, ``{Advanced LIGO},'' {\em Class. Quant. Grav.}, vol.~32,
  p.~074001, 2015, 1411.4547.

\bibitem{TheVirgo:2014hva}
F.~Acernese {\em et~al.}, ``{Advanced Virgo: a second-generation
  interferometric gravitational wave detector},'' {\em Class. Quant. Grav.},
  vol.~32, no.~2, p.~024001, 2015, 1408.3978.

\bibitem{Baker:2019nia}
J.~Baker {\em et~al.}, ``{The Laser Interferometer Space Antenna: Unveiling the
  Millihertz Gravitational Wave Sky},'' {\em preprint (arxiv: 1907.06482)}, Jul
  2019, 1907.06482.

\bibitem{Seto:2001qf}
N.~Seto, S.~Kawamura, and T.~Nakamura, ``{Possibility of direct measurement of
  the acceleration of the universe using 0.1-Hz band laser interferometer
  gravitational wave antenna in space},'' {\em Phys. Rev. Lett.}, vol.~87,
  p.~221103, 2001, astro-ph/0108011.

\bibitem{Harry:2006fi}
G.~Harry, P.~Fritschel, D.~Shaddock, W.~Folkner, and E.~Phinney, ``{Laser
  interferometry for the big bang observer},'' {\em Class. Quant. Grav.},
  vol.~23, pp.~4887--4894, 2006.
\newblock [Erratum: Class.Quant.Grav. 23, 7361 (2006)].

\bibitem{Pierce:2018xmy}
A.~Pierce, K.~Riles, and Y.~Zhao, ``{Searching for Dark Photon Dark Matter with
  Gravitational Wave Detectors},'' {\em Phys. Rev. Lett.}, vol.~121, no.~6,
  p.~061102, 2018, 1801.10161.

\bibitem{Miller:2020vsl}
A.~L. Miller {\em et~al.}, ``{Adapting a semi-coherent method to directly
  detect dark photon dark matter interacting with gravitational-wave
  interferometers},'' {\em preprint (arxiv: 2010.01925)}, Oct 2020, 2010.01925.

\bibitem{Arvanitaki:2014faa}
A.~Arvanitaki, J.~Huang, and K.~Van~Tilburg, ``{Searching for dilaton dark
  matter with atomic clocks},'' {\em Phys. Rev. D}, vol.~91, no.~1, p.~015015,
  2015, 1405.2925.

\bibitem{Morisaki:2018htj}
S.~Morisaki and T.~Suyama, ``{Detectability of ultralight scalar field dark
  matter with gravitational-wave detectors},'' {\em Phys. Rev. D}, vol.~100,
  no.~12, p.~123512, 2019, 1811.05003.

\bibitem{Grote:2019uvn}
H.~Grote and Y.~V. Stadnik, ``{Novel signatures of dark matter in
  laser-interferometric gravitational-wave detectors},'' {\em Phys. Rev. Res.},
  vol.~1, no.~3, p.~033187, 2019, 1906.06193.

\bibitem{Michimura:2020vxn}
Y.~Michimura, T.~Fujita, S.~Morisaki, H.~Nakatsuka, and I.~Obata, ``{Ultralight
  vector dark matter search with auxiliary length channels of gravitational
  wave detectors},'' {\em Phys. Rev. D}, vol.~102, no.~10, p.~102001, 2020,
  2008.02482.

\bibitem{Aoki:2016kwl}
A.~Aoki and J.~Soda, ``{Detecting ultralight axion dark matter wind with laser
  interferometers},'' {\em Int. J. Mod. Phys. D}, vol.~26, no.~07, p.~1750063,
  2016, 1608.05933.

\bibitem{Piffl:2014mfa}
T.~Piffl {\em et~al.}, ``{Constraining the Galaxy's dark halo with RAVE
  stars},'' {\em Mon. Not. Roy. Astron. Soc.}, vol.~445, no.~3, pp.~3133--3151,
  2014, 1406.4130.

\bibitem{Evans:2018bqy}
N.~W. Evans, C.~A. O'Hare, and C.~McCabe, ``{Refinement of the standard halo
  model for dark matter searches in light of the Gaia Sausage},'' {\em Phys.
  Rev. D}, vol.~99, no.~2, p.~023012, 2019, 1810.11468.

\bibitem{2015ApJ...814...13M}
C.~F. {McKee}, A.~{Parravano}, and D.~J. {Hollenbach}, ``{Stars, Gas, and Dark
  Matter in the Solar Neighborhood},'' {\em Astrophys. J}, vol.~814, p.~13,
  Nov. 2015, 1509.05334.

\bibitem{Maggiore:1900zz}
M.~Maggiore, {\em {Gravitational Waves. Vol. 1: Theory and Experiments}}.
\newblock Oxford Master Series in Physics, Oxford University Press, 2007.

\bibitem{Armaleo:2020yml}
J.~M. Armaleo, D.~L\'opez~Nacir, and F.~R. Urban, ``{Pulsar timing array
  constraints on Spin-2 ULDM},'' {\em JCAP}, vol.~09, p.~031, 2020, 2005.03731.

\bibitem{Akrami:2015qga}
Y.~Akrami, S.~F. Hassan, F.~K\"onnig, A.~Schmidt-May, and A.~R. Solomon,
  ``{Bimetric gravity is cosmologically viable},'' {\em Phys. Lett. B},
  vol.~748, pp.~37--44, 2015, 1503.07521.

\bibitem{Armaleo:2019gil}
J.~M. Armaleo, D.~L\'opez~Nacir, and F.~R. Urban, ``{Binary Pulsars as probes
  for Spin-2 Ultralight Dark Matter},'' {\em JCAP}, vol.~2001, no.~01, p.~053,
  2020, 1909.13814.

\bibitem{Babichev:2016bxi}
E.~Babichev, L.~Marzola, M.~Raidal, A.~Schmidt-May, F.~Urban, H.~Veerm\"ae, and
  M.~von Strauss, ``{Heavy spin-2 Dark Matter},'' {\em JCAP}, vol.~09, p.~016,
  2016, 1607.03497.

\bibitem{Murata:2014nra}
J.~Murata and S.~Tanaka, ``{A review of short-range gravity experiments in the
  LHC era},'' {\em Class. Quant. Grav.}, vol.~32, no.~3, p.~033001, 2015,
  1408.3588.

\bibitem{Sereno:2006mw}
M.~Sereno and P.~Jetzer, ``{Dark matter vs. modifications of the gravitational
  inverse-square law. Results from planetary motion in the solar system},''
  {\em Mon. Not. Roy. Astron. Soc.}, vol.~371, pp.~626--632, 2006,
  astro-ph/0606197.

\bibitem{Frasca:2005ey}
S.~Frasca, P.~Astone, and C.~Palomba, ``{Evaluation of sensitivity and
  computing power for the Virgo hierarchical search for periodic sources},''
  {\em Class. Quant. Grav.}, vol.~22, pp.~S1013--S1019, 2005.

\bibitem{DAntonio:2018sff}
S.~D'Antonio {\em et~al.}, ``{Semicoherent analysis method to search for
  continuous gravitational waves emitted by ultralight boson clouds around
  spinning black holes},'' {\em Phys. Rev. D}, vol.~98, no.~10, p.~103017,
  2018, 1809.07202.

\bibitem{Brady:1998nj}
P.~R. Brady and T.~Creighton, ``{Searching for periodic sources with LIGO. 2.
  Hierarchical searches},'' {\em Phys. Rev. D}, vol.~61, p.~082001, 2000,
  gr-qc/9812014.

\bibitem{Krishnan:2004sv}
B.~Krishnan, A.~M. Sintes, M.~A. Papa, B.~F. Schutz, S.~Frasca, and C.~Palomba,
  ``{The Hough transform search for continuous gravitational waves},'' {\em
  Phys. Rev. D}, vol.~70, p.~082001, 2004, gr-qc/0407001.

\bibitem{Antonucci:2008jp}
F.~Antonucci, P.~Astone, S.~D'Antonio, S.~Frasca, and C.~Palomba, ``{Detection
  of periodic gravitational wave sources by Hough transform in the f versus
  $\dot{f}$ plane},'' {\em Class. Quant. Grav.}, vol.~25, p.~184015, 2008,
  0807.5065.

\bibitem{PhysRevD.90.042002}
P.~Astone, A.~Colla, S.~D'Antonio, S.~Frasca, and C.~Palomba, ``Method for
  all-sky searches of continuous gravitational wave signals using the
  frequency-hough transform,'' {\em Phys. Rev. D}, vol.~90, p.~042002, Aug
  2014.

\bibitem{Hild:2010id}
S.~Hild {\em et~al.}, ``{Sensitivity Studies for Third-Generation Gravitational
  Wave Observatories},'' {\em Class. Quant. Grav.}, vol.~28, p.~094013, 2011,
  1012.0908.

\bibitem{Evans:2016mbw}
B.~P. Abbott {\em et~al.}, ``{Exploring the Sensitivity of Next Generation
  Gravitational Wave Detectors},'' {\em Class. Quant. Grav.}, vol.~34, no.~4,
  p.~044001, 2017, 1607.08697.

\bibitem{Schmitz:2020syl}
K.~Schmitz, ``{New Sensitivity Curves for Gravitational-Wave Signals from
  Cosmological Phase Transitions},'' {\em JHEP}, vol.~01, p.~097, 2021,
  2002.04615.

\bibitem{Zhang:2019oet}
C.~Zhang, Q.~Gao, Y.~Gong, D.~Liang, A.~J. Weinstein, and C.~Zhang,
  ``{Frequency response of time-delay interferometry for space-based
  gravitational wave antenna},'' {\em Phys. Rev. D}, vol.~100, no.~Jun,
  p.~064033, 2019, 1906.10901.

\bibitem{Stott:2020gjj}
M.~J. Stott, ``{Ultralight Bosonic Field Mass Bounds from Astrophysical Black
  Hole Spin},'' {\em preprint (arxiv: 2009.07206)}, Sep 2020, 2009.07206.

\bibitem{Brito:2020lup}
R.~Brito, S.~Grillo, and P.~Pani, ``{Black Hole Superradiant Instability from
  Ultralight Spin-2 Fields},'' {\em Phys. Rev. Lett.}, vol.~124, no.~21,
  p.~211101, 2020, 2002.04055.

\bibitem{Lee:2008}
K.~J. {Lee}, F.~A. {Jenet}, and R.~H. {Price}, ``{Pulsar Timing as a Probe of
  Non-Einsteinian Polarizations of Gravitational Waves},'' {\em Astrophys.\
  J.}, vol.~685, pp.~1304--1319, Oct. 2008.

\end{thebibliography}
\end{document}